\documentclass[prd,twocolumn,superscriptaddress,floatfix,amsmath,amssymb]{revtex4-1}
\usepackage[english]{babel}
\usepackage{graphicx}% Include figure files
\usepackage{dcolumn}% Align table columns on decimal point
\usepackage{bm}% bold math
\usepackage{verbatim}
\usepackage{mathrsfs}
\usepackage{cancel}
\usepackage{epstopdf}
\usepackage{color}
\usepackage[usenames,dvipsnames]{pstricks}
\usepackage{epsfig}
\usepackage{pst-grad} % For gradients
\usepackage{pst-plot} % For axes

\usepackage{hyperref}

\newcommand{\proj}[2]{\left| {#1} \right\rangle\!\left\langle {#2} \right|}

\newcommand{\nn}{\nonumber}
\newcommand{\be}{\begin{equation}}
\newcommand{\ee}{\end{equation}}
\newcommand{\bea}{\begin{eqnarray}}
\newcommand{\eea}{\end{eqnarray}}

\def\slashchar#1{\setbox0=\hbox{$#1$} % set a box for #1
\dimen0=\wd0 % and get its size
\setbox1=\hbox{/} \dimen1=\wd1 % get size of /
\ifdim\dimen0>\dimen1 % #1 is bigger
\rlap{\hbox to \dimen0{\hfil/\hfil}} % so center / in box
#1 % and print #1
\else % / is bigger
\rlap{\hbox to \dimen1{\hfil$#1$\hfil}} % so center #1
/ % and print /
\fi}

\begin{document}

\title{Cavities in curved spacetimes: the response of particle detectors}
\author{Aida Ahmadzadegan}
\email{aida.ahmad@uwaterloo.ca}
\affiliation{Department of Physics and Astronomy, University of Waterloo, Waterloo, Ontario N2L 3G1, Canada}
\author{Eduardo Mart\'{i}n-Mart\'{i}nez}
\email{emartinm@uwaterloo.ca}
\affiliation{Institute for Quantum Computing, University of Waterloo, Waterloo, Ontario, N2L 3G1, Canada}
\affiliation{Department of Applied Math, University of Waterloo, Waterloo, Ontario, N2L 3G1, Canada}
\affiliation{Perimeter Institute for Theoretical Physics, 31 Caroline St N, Waterloo, Ontario, N2L 2Y5, Canada}
\author{Robert B. Mann}
\email{rbmann@uwaterloo.ca}
\affiliation{Department of Physics and Astronomy, University of Waterloo, Waterloo, Ontario N2L 3G1, Canada}
\begin{abstract}

We introduce a method to compute a particle detector transition probability in spacetime regions of general curved spacetimes provided that the curvature is not above a maximum threshold. In particular we use this method to compare  the response of two detectors, one in a spherically symmetric gravitational field and the other one in Rindler spacetime to compare the Unruh and Hawking effects: We study the vacuum response of a detector freely falling through a stationary cavity in a Schwarzschild background as compared to the response of an equivalently accelerated detector traveling through an inertial cavity in the absence of curvature. We find that as we set the cavity at increasingly further radii from the black hole, the thermal radiation measured by the detector approaches the quantity recorded by the detector in Rindler background showing in which way and at what scales the equivalence principle is recovered in the Hawking-Unruh effect. I.e. when the Hawking effect in a Schwarzschild background becomes equivalent to the Unruh effect in Rindler spacetime.

\end{abstract}

\maketitle

\section{Introduction}\label{Intro}

%\textcolor[rgb]{0.2,0.8,0.2}{Quantum field theory in curved spacetime has been a commonplace to study the interaction between gravitation and quantum field theory and so far, it has provided us with a profound insights and intuitions into the nature of quantum gravity.}

One of the renowned predictions of Quantum field theory in curved spacetime is called the `Unruh effect' \cite{Unruh1976,Unruh2005,Crispino,Birrell1984,Wald1994,Unruh-Wald}. It was suggested by Unruh that uniformly accelerated observers in Minkowski spacetime detect a thermal distribution of particles (with a temperature proportional to their proper acceleration) when probing the vacuum state for an inertial observer.  It is indeed a challenge to detect this effect using present technology, as it involves measuring low temperatures with `thermometric probes' that move with extremely high accelerations.
%This lack of experimental confirmation has brought the existence of this effect to debates \cite{sceptic}.
%A number of very strong assumptions have to be made in order to justify the observation of a thermal bath by an accelerated observer in the Minkowski vacuum. For example, the infinite amount of energy required to sustain the eternal Rindler trajectory, or the difficulties that arise in defining a Minkowski vacuum when boundary conditions are specified on the scalar field on a manifold \cite{Eric B.}.

To model the response of an accelerated probe measuring the quantum field, it is commonplace to use the so-called Unruh-DeWitt detector (UdW) \cite{DeWitt,Birrell1984} which is an idealized model of a real particle detector that still encompasses all the fundamental features of the light-matter interaction when there is no angular momentum exchange involved \cite{Wavepackets}. This model consists of a two level quantum system that  couples in pointlike manner to a scalar field along its worldline.   Notice that, following the usual nomenclature  in the literature, we refer to  the Unruh-DeWitt detector as `particle detector' \cite{Birrell1984,LeeThesis}, and that we will focus in this paper on the response of an Unruh-Dewitt detector rather than in performing a fundamental analysis of the particle content of the field. 

The response of a particle detector depends on the state of the field, the structure of spacetime and the detector's trajectory. Regarding different types of trajectories, a number of studies have been done in Minkowski spacetime and a variety of scenarios have been explored \cite{Satz2006,Obadia,Hodgkinsonclick}. For example, the transition rate of UdW detector coupled to a massless scalar field in Minkowski spacetime, regularized by the spatial profile, was analyzed in \cite{Satz2006}. 

However,  one faces technical difficulties when in studying the response of detectors undergoing non-trivial trajectories in curved backgrounds. Notwithstanding simple two dimensional cases  \cite{Singleton}, in curved spacetimes  the identification of the correct vacuum state of the theory, which provides a physical interpretation of the results, is difficult because of the lack of a global timelike Killing vector. Even when we can identify the relevant vacuum, the obtention of the Wightman function and its appropriate regularization in general backgrounds can also pose a challenging problem even for the simplest cases \cite{Hodgkinson2012}.

%For example, if one tries to measure the probability of transition of the detector at the time in which the interaction is still going on, the response function of the detector would be no longer well defined as the switching function has a sharp cut-off at a singularity of the Wightman distribution \cite{transitioncurvedLouko}....

Given these difficulties it is interesting to explore some approximate regimes where it may be possible to find the response of a particle detector without running into  the severe technical complications suffered by exact methods. In principle, approaches using cavity quantum field theory have been explored in order to answer questions regarding the particle content of the vacuum state of a field from the perspective of different observers 
%(Mart\'{i}n-Mart\'{i}nez et al. 
\cite{AasenPRL,Brown2012,Fuenetesevolution}. In those approaches the treatment gets greatly simplified by the fact that the  cavity field gets isolated from the rest of the free field in the spacetime and so an IR-cutoff is built into the theory. Futhermore, the problem of studying the response of particle detectors in optical cavities in relativistic regimes is a problem of intrinsic interest \cite{Brown2012,Fuenetesevolution}; a recent result showed that  an accelerated detector inside  a cavity does very approximately  thermalize to a temperature proportional to its acceleration \cite{Wilson2013}.

Here we investigate the difference in the response of a detector when freely falling through a stationary cavity subjected to a spherically symmetric gravitational field as compared to the response of an equivalently accelerated detector that traverses an inertial cavity in the absence of curvature. One might expect (via the equivalence principle) both responses to be similar if the cavity is small compared to the distance from the source of the gravitational field. However, as we will see, as the cavity is placed closer to the region of strong gravity the detector responses increasingly differ. 

With the help of some approximations that are applicable in the cavity scenario, we will be able to characterize the transition probability of particle detectors in a Schwarzschild background, circumventing the complexity involved in the calculation of the Wightman function in such scenarios \cite{Louko2008}.

To this end, the outline of our  paper is as follows. In sec. \ref{set}, we introduce the Physical model of our system including the methodology for investigating our cavity scenario. Sec. \ref{transition}, contains a discussion of our results.  We finish with sec. \ref{concl} with concluding remarks.

\section{The setting}\label{set}

In order to investigate our method, we consider the following scenario to calculate and analyze the excitation probability of an UdW detector in the cavity. We consider a detector to be a two-level pointlike quantum system, which is a reasonable approximation for atomic-based particle detectors \cite{Wavepackets}.  We will be working with three different coordinate systems. One is the coordinate system of the stationary observer at infinity, $(r,t)$, the second one is the local coordinate system of the outermost wall of the cavity, $(r',t')$ which is sitting at the radius $(r=R)$ and the third one is the proper frame of the detector, whose proper time we denote as $\tau$. 

The  proper frame $(r',t')$ of the outermost wall is related to the asymptotically stationary frame $(r,t)$ by means of the following relationships: 
\bea
r'&=&\left(1-\frac{2m}{R}\right)^{-\frac{1}{2}}(R-r),\\
t'&=&\left(1-\frac{2m}{R}\right)^{\frac{1}{2}}t.
\eea

\begin{figure}[htp]
	\includegraphics[width=0.5\textwidth]{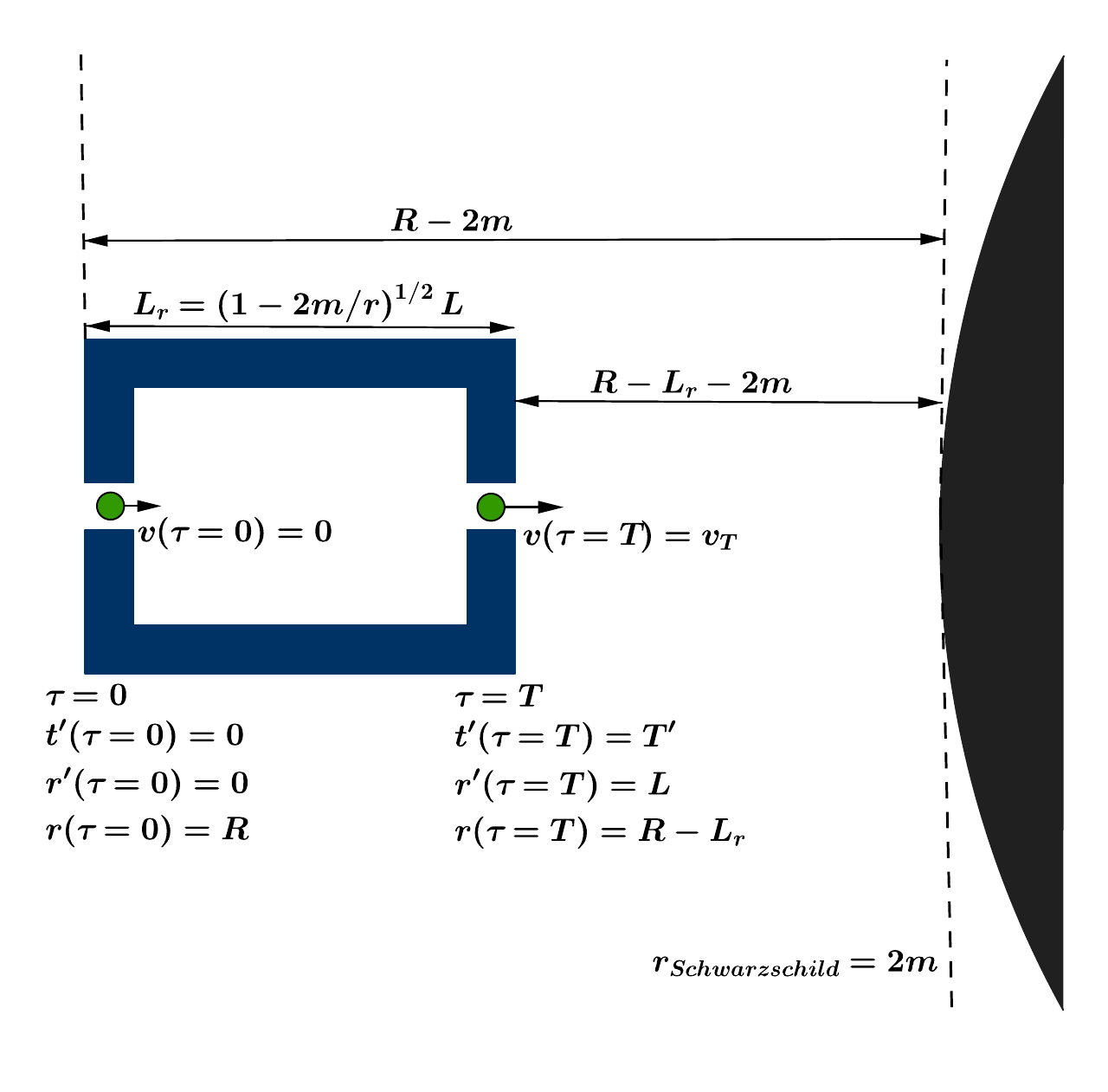}
  \caption{Scheme of a detector going through a cavity prepared in the vacuum state, in a curved background. The cavity, of proper length $L$ (Length $L_r$ in the asymptotically flat frame) is located at the arbitrary radius $R$ in the asymptotically flat frame $(r,t)$. The detector with zero initial velocity is falling through the cavity and it spends an amount of proper time of $T$ (or equivalently $T'$ from the cavity frame $(r',t')$) to travel through the cavity, exiting  it with final proper speed $v_T$.}
  \label{setup}
\end{figure}

Now, we want to parametrize the trajectory of a free-falling detector, which starts   from rest at the beginning of the cavity ($r=R\Rightarrow r'=0$), in terms of its own proper time $\tau$. We would like to write the worldline of the detector in terms of a parametric curve in the cavity's frame coordinates, i.e. we want the detector's worldline $(r'[\tau],t'[\tau])$. The parametrization of the detector's trajectory in this frame is given by
\begin{align}
&r'\left[\theta(\tau)\right]=\left(1-\frac{2m}{R}\right)^{-\frac{1}{2}} R\sin^2 \left(\frac{\theta(\tau)}{2}\right),\\
&t'\left[\theta(\tau)\right]=\left(1-\frac{2m}{R}\right)\left(\frac{R^3}{2m}\right)^{\frac{1}{2}}\bigg[\frac{1}{2} \Big(\theta(\tau)+\sin \left[\theta(\tau)\right]\Big)\nn\\&+\frac{2m}{R}\theta(\tau)\bigg]+\left(1-\frac{2m}{R}\right)^{\frac{1}{2}} 2m \log \left[\frac{\tan \frac{\theta_H}{2}+\tan \frac{\theta(\tau)}{2}}{\tan \frac{\theta_H}{2}-\tan \frac{\theta(\tau)}{2}}\right]\label{time},
\end{align}
where  $\theta$ is the following function of the proper time of the detector $\tau$: 
\bea
\theta(\tau)=2\arccos \left(\frac{r(\tau)}{R}\right)^{\frac{1}{2}}.
\eea
and  $\theta_H$ is the value of $\theta$ at the horizon \cite{MathBH}. We depict the scheme for this setting  in Fig. \ref{setup}.

The size of the cavity will be considered small enough so we can ignore any tidal effects.   In this fashion we will assume that the reference frame of the outermost wall of the cavity will very approximately be the same frame as for all the rest of the points in the whole cavity. We will discuss the validity of this approximation later on; taking it as valid, the stationary cavity placed at $r=R$ has length $L$ in its local coordinate system. Consequently,  the length of the cavity as measured by a stationary observer at infinity, $L_r$, is related to $L$ via $L_r=\left(1-2m/r\right)^{1/2}L$.

Now one would expect that a detector falling through the cavity, even if field and detector are in the ground state, would experience a  Hawking-effect-like response. There are mainly two contributions to the distinct response of the detector in this regime: First, since the detector is freely falling, its proper time is different at each point in  the cavity. Second, the solution to the Klein-Gordon equation would be different from the usual stationary waves in a flat-spacetime Dirichlet cavity.

In our model, we will carry out the following  `quasi-local' approximation. If the cavity is small enough and it is far enough from the strong gravity region, we can assume that the solutions of the Klein-Gordon (KG) equation inside the cavity can be very well approximated by plane waves in the locally flat tangent spacetime (corrections due to the effects of curvature can be incorporated via the Riemann normal coordinate expansion \cite{Bunch:1979uk}).  While moving through the cavity, the proper time of the detector will still experience a gravitational redshift, which will be responsible for its thermal response. We expect that departures in the KG solutions with respect to the flat spacetime scenario will introduce sub-leading corrections in the appropriate  regimes; it is these corrections that we are neglecting.

To check the range of validity of this estimation we proceed as follows: KG equation in a Schwarzschild background has the following form where $r$ and $t$ are the coordinates of the stationary observer in the asymptotically flat region,
\bea
\Bigg[\partial^2_{\text t}-\frac{1}{r^2}\left(1-\frac{2m}{r}\right)\bigg(\partial_{\text r}(r^2-2mr)\partial_{\text r}+\Delta_S\bigg)\Bigg]\psi(t,r)=0\nn
\eea
where $\Delta_S$ is the   Laplacian on $S^2$.  This expression can be written in a similar form as in flat spacetime if we write it in terms of the Regge-Wheeler coordinate
\bea
r_{\ast}=r+2m \ln\left(\frac{r}{2m}-1\right),
\eea
as
\bea\label{KGRW}
\left[\partial^2_{\text t}-\partial^2_*+V(r)\right]\psi(t,r)=0
\eea
%\bea
%\frac{d}{dr_{\ast}}=\left(1-\frac{2m}{r}\right)\frac{d}{dr}
%\eea
where
\bea 
V(r)=\left(1-\frac{2m}{r}\right)\left(\frac{2m}{r^3}-\frac{\Delta_S}{r^2}\right)\nn
\eea
In two regimes $V(r)$ approaches zero: One is close to the horizon where $r \rightarrow 2m$ and the other one is far away from the horizon where $r \rightarrow \infty$. In these two ranges, the form of KG equation in the Regge-Wheeler frame would be effectively of the same form as the flat spacetime equation. Nevertheless, the KG equation in the cavity frame $(r',t')$ (where we carry out the field quantization) will not be of that form. We need to find an  estimator that tells us how precisely we can approximate the KG equation in a Schwarzschild background with the one in a locally flat background associated with the rest frame of the cavity. If the length of the cavity is small enough,  the ratio between the length of the cavity in its own reference frame $(\Delta r'=L)$ and the length  of the cavity in the Regge-Wheeler frame $(\Delta r_*=L_*)$  provides a physically meaningful estimator of the validity of the quasi-local approximation. Given the relationship between $r_*$ and $r'$ for the radially ingoing detector
\be
r_{\ast}\!=\!-R+\left(1-\frac{2m}{R}\right)^{\frac{1}{2}} \!\!r'-2m \ln \!\left[\frac{R-\left(1-\frac{2m}{R}\right)^{\frac{1}{2}} r'}{2m}\!-\!1 \right]\!,
\ee
the estimator takes the following form
\be\label{estimator1}
\frac{L_\ast}{L}=\frac{\Delta r_{\ast}}{\Delta r'}=\left(1-\frac{2m}{R}\right)^\frac{1}{2}-\frac{2m}{L}\ln \left[\frac{(R^2-2mR)^\frac{1}{2}}{(R^2-2mR)^\frac{1}{2}-L}\right].
\ee

\begin{figure}[htp]
	\includegraphics[width=0.48\textwidth]{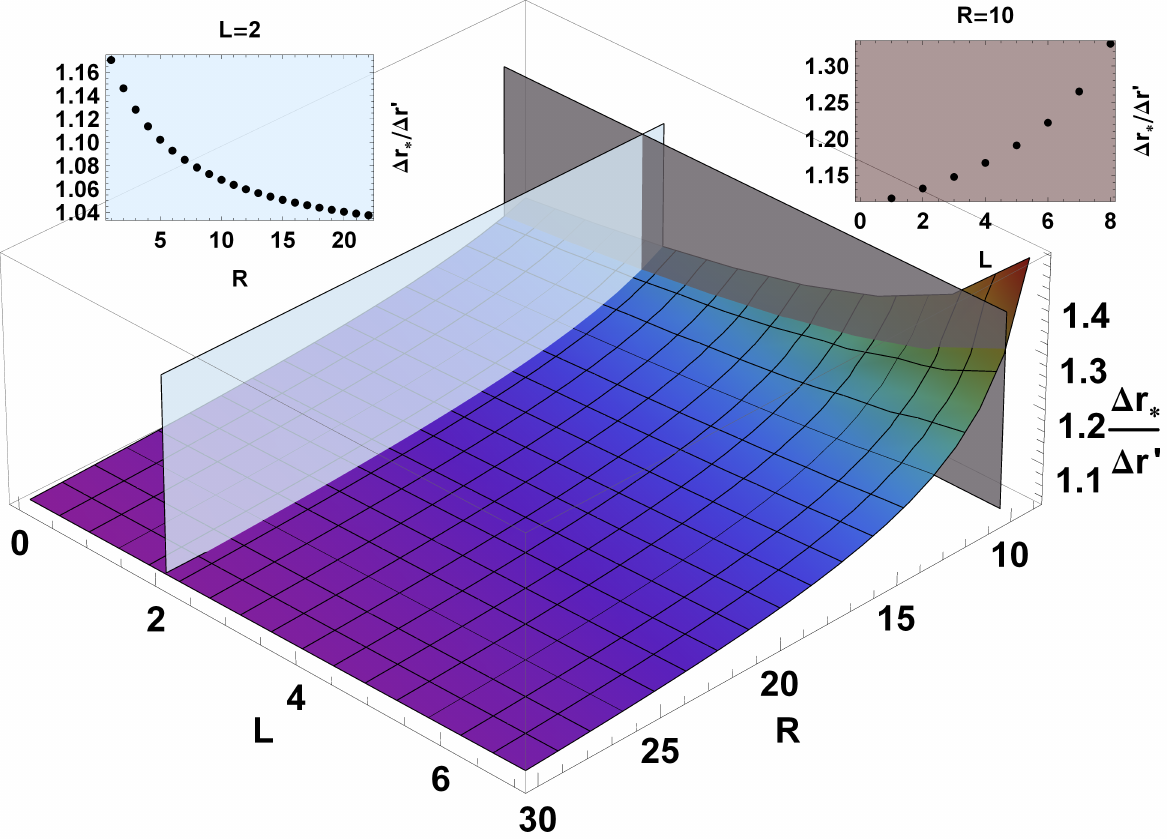}
  \caption{Estimator of the validity of the model: The closer to 1, the better the approximation. We see that, as expected, when we are far away from the horizon and when we consider small cavities, $\Delta r_{\ast}/\Delta r'$ approaches one. The inset located at the top right shows the behavior of the ratio vs. $R$ for a fixed value of $L=2$ and the top left one shows how ratio changes with $L$ when the cavity is at $R=10$ (all magnitudes are expressed in units of black hole mass).}
  \label{estimator}
\end{figure}

Figure \ref{estimator} illustrates how this quantity changes as a function of $R$, the distance from the black hole and $L$, the length of the cavity.  We see that $\Delta r_{\ast}/\Delta r'$ increases with both increasing cavity size and decreasing proximity to the Schwarzschild radius. Our approximation works well  for the small cavities in the vicinity of the black hole. As we will see, the threshold size of the cavity is equal to the size of the black hole.

 Note again that the estimator is reliable when we are very close to the event horizon, or far away from it, as discussed above. Nevertheless, and as an interesting remark, in a 1+1 dimensional Schwarzschild background, it is easy to prove that the Klein-Gordon equation in the Regge-Wheeler frame has the same form as \eqref{KGRW} but with $V(r)=0, \forall r$. In this case  the estimator is reliable for any position of the cavity.

The Hamiltonian describing our system consists of three terms: $\hat{H}^{(\text{d})}_{\text{free}}$, the free Hamiltonian of the detector, $\hat{H}^{(\text{f})}_{\text{free}}$, the free Hamiltonian of the field, and the field-detector interaction Hamiltonian $\hat{H}_{\text{int}}$:
\bea
\hat{H}=\hat{H}^{(\text{d})}_{\text{free}}+\hat{H}^{(\text{f})}_{\text{free}}+\hat{H}_{\text{int}}.
\eea
We will model the detector-field interaction with the well-known Unruh-DeWitt model \cite{DeWitt,Louko2008}
\bea
\hat{H}_{\text{int}}= \lambda\chi(\tau) \hat{\mu}(\tau) \hat{\phi}\left[r'(\tau)\right],
\eea
where the constant $\lambda$ is the coupling strength, $\chi(\tau)$ is the {\it switching function} or {\it time window function} controlling the smoothness of switching the interaction on and off.  $\hat{\mu}(\tau)$ is the monopole moment of the detector and $\hat{\phi}\left[r'(\tau)\right]$ is the massless scalar field to which the detector is coupling. We consider the coupling constant to be a small parameter so we can work with  perturbation theory to   second order in $\lambda$. In our setting, the switching function is nonvanishing only during the time the atom spends in the cavity, i.e., $\chi(\tau)=1$ during $0\le \tau \le T$. The monopole moment operator of the detector has the usual form in the interaction picture, % smoothly to make the quantum noise effects negligible comparing with the response of the detector while accelerating \cite{Brownedu}. 
\bea
\hat{\mu}_{\text{d}}(\tau)=(\sigma^{+}e^{i\Omega_{\text{d}} \tau}+\sigma^{-}e^{-i\Omega_{\text{d}} \tau}),
\eea
in which, $\Omega_{\text{d}}$ is the proper energy gap between the ground state, $\left| g \right\rangle$ and the excited state, $\left| e \right\rangle$ of the detector and $\sigma^{-}$ and $\sigma^{+}$ are Ladder operators.

Expanding the field  in terms of an orthonormal set of solutions inside the cavity yields the Hamiltonian in the interaction picture \cite{Robort}
\be\label{hamilto}
\hat{H}_{\text{int}}=\lambda \frac{d\tau}{dt}\sum^{\infty}_{n=1}\frac{\hat{\mu}_{\text{d}}(t)}{\sqrt{\omega_n L}}\big(\hat{a}^{\dagger}_{n}u_n\left[r'(\tau),t'(\tau) \right]+\hat{a}_{n}u^{*}_n\left[r'(\tau),t'(\tau)\right]\big)
\ee
%where the normalization factor of $1/\sqrt{\omega_n L}$ comes from the Klein-Gordon inner product of scalar fields and $\hat{a}^{\dagger}_{n}$ and $\hat{a}_{n}$ are the creation and annihilation operators of the field modes, respectively \cite{Birrell},
We will consider  Dirichlet (reflective) boundary conditions for our cavity, 
(see Fig. \ref{setup})
\bea
\phi\left[0,t'\right]=\phi \left[L,t'\right]=0
\eea
and so under our quasi-local approximation the field modes take the form of the stationary waves
\bea
u_n\left[r'(\tau),t'(\tau)\right]=e^{i\omega_n t'(\tau)}\sin \left[k_n r'(\tau)\right].
\eea
Note that $\omega_{n}=\left|k_n\right|=n \pi/L$, and $r'(\tau)$ and $t'(\tau)$ (given in equations \eqref{time} respectively)  parameterize the trajectory of the detector freely falling from a cavity whose first wall is located at $r=R$ in the frame $(r,t)$, proper to a stationary observer at infinity.

We want to characterize the vacuum response of a particle detector undergoing the trajectory \eqref{time}. For our purposes, we prepare the detector in its ground state and the cavity in the vacuum state
\bea \label{initial}
\rho_0=\proj{g}{g} \otimes \proj{0}{0}
\eea
To proceed, we let this detector start free falling through the cavity as shown in Fig. \ref{setup}. The detector spends an amount of proper time $T$ inside the cavity. The time evolution of the system is governed by the interaction Hamiltonian \eqref{hamilto} in the time interval $0<\tau<T$, whereas for the detector  in the cavity it is given by the Dyson's perturbative expansion 
\begin{align}\label{Dyson}
\hat{U}(T,0)&=\openone\underbrace{-i\int^{T}_{0}d\tau \hat{H}_{\text{int}}(\tau)}_{\hat{U}^{(1)}}\nn \\
&\underbrace{+(-i)^2\int^{T}_{0}d\tau \int^{\tau}_{0}d\tau' \hat{H}_{\text{int}}(\tau)\hat{H}_{\text{int}}(\tau')}_{\hat{U}^{(2)}}+ ...\nn\\
&\underbrace{+(-i)^n\int^{T}_{0}d\tau ... \int^{\tau^{(n-1)}}_{0}d\tau^{(n)} \hat{H}_{\text{int}}(\tau) ... \hat{H}_{\text{int}}(\tau^{(n)})}_{\hat{U}^{(n)}}
\end{align}
for the time evolution operator. 

In this model, we assume that the detector is weakly coupled to the field and keep the terms in the expansion \eqref{Dyson} up to the second order of perturbation in $\lambda$. The system's density matrix at a time $T$ would be evaluated as \cite{AasenPRL}
\be
\rho_{T}\!=\!\big[\openone+\hat{U}^{(1)}+\hat{U}^{(2)}+\mathcal{O}(\lambda^3)\big]\rho_0\big[\openone+\hat{U}^{(1)}+\hat{U}^{(2)}+\mathcal{O}(\lambda^3)\big]^{\dagger}
\ee
which we write as
\bea\label{finaldensity}
\rho_{T}=\rho_{0}+\rho^{(1)}_{T}+\rho^{(2)}_{T}+\mathcal{O}(\lambda^3),
\eea
where
\bea
\rho^{(0)}_{T}&=&\rho_0,\\
\rho^{(1)}_{T}&=&\hat{U}^{(1)}\rho_0+\rho_0 \hat{U}^{(1)\dagger},\\
\rho^{(2)}_{T}&=&\hat{U}^{(1)}\rho_0 \hat{U}^{(1)\dagger}+\hat{U}^{(2)}\rho_0+\rho_0 \hat{U}^{(2)\dagger}.
\eea
Using \eqref{hamilto} and \eqref{Dyson}, the first order term of the perturbative expansion takes the form
\begin{align}
\hat{U}^{(1)}=\frac{\lambda}{i}\sum^{\infty}_{n=1}&\big[\sigma^{+} a^{\dagger}_{n}I_{+,n}+\sigma^{-} a_{n}I^{\ast}_{+,n}\nn\\
&+\sigma^{-} a^{\dagger}_{n}I_{-,n}+\sigma^{+}a_{n}I^{\ast}_{-,n}\big],
\end{align}
in which
\bea
I_{\pm,n}=\int^{T}_0 d\tau~e^{i\left[\pm \Omega_{\text d} \tau+\omega_nt'(\tau)\right]} \sin\left[k_n r'(\tau)\right].
\eea

%Note that the phases in the integrand are time dependent because of $r'(\tau)$ and $t'(\tau)$.
To compute the density matrix for the detector, $\rho_{T}^{(\text{d})}$, we need to take the partial trace over the field degrees of freedom  \cite{AasenPRL}. The first order contribution to the probability of transition vanishes, so the leading contribution comes from   second order in the coupling strength, $\lambda$. The final form of the detector density matrix is
\begin{align}\label{densitytrace}
\rho_{T,(\text{d})}&=\text{Tr}_{(\text{f})}\!\left[\rho_{0}+\hat{U}^{(1)}\rho_0\hat{U}^{(1)\dagger}\!+\hat{U}^{(2)}\rho_0+\rho_0\hat{U}^{(2)\dagger}\right]\!\!,
\end{align}
which yields
\bea
\rho_{T,(\text{d})}=\text{Tr}_{\text{f}}\, \rho_T=\left[\begin{array}{ll}
1-P_2&0\\
0&P_1\end{array}\right]
\eea
where  $P_1$ and $P_2$  are given by
\begin{align}\label{P1}
P_1&=\lambda^2\sum_{n=1}^{\infty}\left|I_{+,n}\right|^2,\\
P_2&= \lambda^2 \sum^{\infty}_{n=1} 2\text{Re}(J_n),
\end{align}
and where
\begin{align}
J_n&=\int^{T}_0 d\tau \int^{\tau}_0 d\tau_1 \Big[e^{-i\left[\Omega_{\text{d}} \tau+\omega_nt'(\tau)\right]} \sin \left[k_nr'(\tau)\right]\nn\\
&\times e^{i\left[\Omega_{\text{d}} \tau_1+\omega_nt'(\tau_1)\right]} \sin \left[k_nr'(\tau_1)\right]\Big].
\end{align}

$P_1$ gives the transition probability of  the detector from the ground state to the first excited state to the leading order in perturbation theory. 
%Since we are working with just one detector, the off-diagonal (correlation) terms are zero so our detector is in a mixed state which means $\rho^2_{(d)}\neq \rho_{(d)}$. $Tr(\rho^2_{(d)})$ can vary between $\frac{1}{d}$ for maximally mixed state to $1$ as in pure state. In our case, we observe that $Tr(\rho^2_{(d)})\approx 0.9999999568$ which is close to one and means that the degree of mixedness of the detector is not high and it is close to the pure state. $P_1$ gives us the excitation probability (transition rate) of the detector anty d $P_2$ ...

 Note that we have decided to compute the probability of transition rather than the transition rate. Given the absence of time translational invariance of our setting, there is no formal or computational advantage in computing the rate over the probability, and both magnitudes contain the same information.

We furthermore note that the transition probability of a suddenly switched detector becomes logarithmically  divergent in the 3+1 dimensional case, but it is finite in lower dimensional scenarios  \cite{Padmanabhan,Louko2008},  effectively rendering our calculation in the relevant radial coordinate (where we assume the cavity is longer) divergence-free. A repetition of the same calculation in 3 spatial dimensions would entail making the switching function of the detector continuous. Alternatively one can compute differences between transition probabilities as in \cite{Louko2008,QuanG}. Since in this article we are comparing the Rindler with the Schwarzschild detector response  the fundamental results reported here would not be modified by these effects.

\section{Transition probability of the detector}\label{transition}

Applying the formalism of section \ref{set}, we proceed to present and compare our results for the response of the detector freely falling in Schwarzschild spacetime to that of an accelerated one in a Minkowski background.

For the Schwarzschild case, we consider a free-falling detector   passing through a cavity. It starts falling with zero initial velocity at at $r=R$, the entrance of the cavity.  We assume the the detector enters the cavity in the ground state   (of the free Hamiltonian) and that the field in the cavity is prepared in the local vacuum state. This set up is shown schematically in Fig. \ref{setup}. The detector gets excited due to the difference between its proper time and the proper time in the cavity frame, which induces an effective time dependence in the interaction Hamiltonian. 

Using equation \eqref{P1}, we find the transition probability. We select an arbitrary value (of $\lambda=0.01$) for the coupling strength and set $\Omega=6\pi/L$ so the detector resonates with the 6th mode of the field in the cavity.   This somewhat arbitrary choice is convenient in that by coupling  the detector to a higher harmonic of the cavity we avoid its decoupling from the cavity field by being taken off-resonance through the blueshift the field modes experience in the detector's frame \cite{Wilson2013}.

\begin{figure}[htp]
	\includegraphics[width=0.48\textwidth]{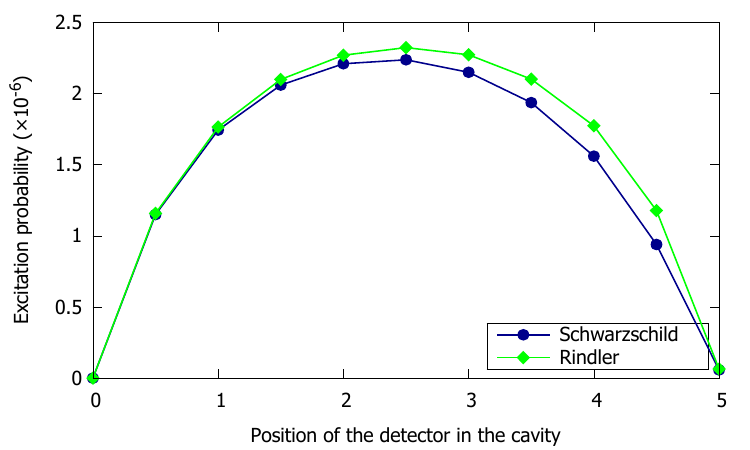}
  \caption{Excitation probability of the detector while it is traveling from the beginning to the end of the cavity of size $L=5$. The green (squared) curve indicates the transition probability in the Rindler background and the blue (circled) curve is for the case of Schwarzschild spacetime. The coupling strength set to be $\lambda=0.01$. }
  \label{cavityprob}
\end{figure}

Now, to compare these results with the Rindler scenario, we set a cavity of the same proper length in a Minkowski background, traversed by an accelerated detector (of proper acceleration $a$ equal to the gravitational field intensity in the spherically symmetric Schwarzschild brackground at a radius $r=R$) and with the same energy gap as above. The detector's worldline, parametrized in terms of its proper time is
\bea
x(\tau)&=&\frac{1}{a}\left(\cosh(a\tau)-1\right),\\
t(\tau)&=&\frac{1}{a}\sinh(a\tau)
\eea
so that the detector is at $x=0$ (the cavity entrance)  at time $t=\tau=0$. By inserting these functions in the interaction Hamiltonian \eqref{hamilto} (substituting $r'$ and $t'$ by $x$ and $t$) and following the same calculation that we did for the Schwarzschild case, we find the transition probability for the detector in Rindler spacetime. The behavior of the response of the accelerated detector while it is passing through the cavity is shown in Fig. \ref{cavityprob} (green squared curve). As noted above, to compare the excitation probability of the detector in Rindler spacetime with those at different radii in the Schwarzschild background, the detector's proper accelerations in the  Rindler scenario will be taken to be
\bea
a=\frac{m}{R^2\left(1-\frac{2m}{R}\right)^\frac{1}{2}}
\eea
so that they are equivalent to the real acceleration measured by the detector at the specific radii considered in the curved background (i.e. the local strength of the gravitational field).  

Plotting all quantities in units of the black hole mass $m$, Fig. \ref{cavityprob} shows the behaviour of the excitation probability of the detector while it is traveling from the beginning to the end of a cavity of size $L=5$ located at radius $R=10$ for both the Schwarzschild and Rindler cases.

\begin{figure}[htp]
	\includegraphics[width=0.48\textwidth]{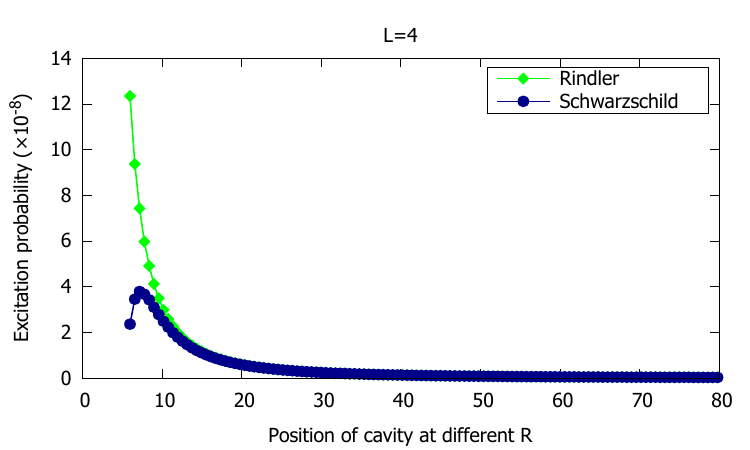}
  \caption{Behavior of the detector transition probability as it starts falling through the cavity of length $L=4$ from different radii $R$ for Schwarzschild (circled curve) and Rindler (squared curve) spacetimes.}
  \label{cavityprobs}
\end{figure}

%When we set the cavity at different radii, the transition probability is changing. 
Figure \ref{cavityprobs} shows how the transition rate changes as we set the cavity at different distances from the black hole for the cavity of fixed length and the consistent change in the atom's acceleration for the Rindler cavity.

\begin{figure*}[htp]
	\includegraphics[width=0.72\textwidth]{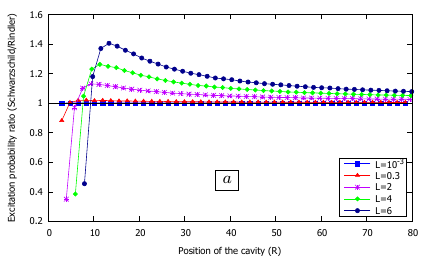}
		\includegraphics[width=0.72\textwidth]{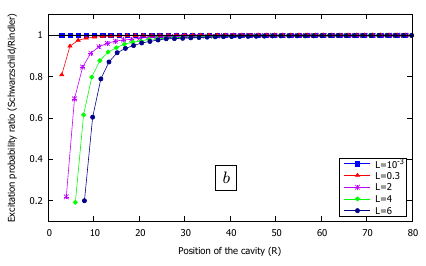}
  \caption{The ratio of the transition probability of an Unruh-DeWitt detector in a Schwarzschild background to the equivalent quantity in Rindler spacetime. Each curve shows the behaviour of the ratio for a different cavity length  as it gets placed at different $R$ from the Schwarzschild black hole and equivalently gets assigned with different constant accelerations equal to the Schwarzschild acceleration at the $a)$ entrance to the cavity and $b)$ middle of the cavity.}
  \label{ratiograph}
	\end{figure*}

To compare our results in the presence and absence of curvature, we present the ratio of the probabilities of  the two scenarios  in Fig. \ref{ratiograph}. Each curve represents the behaviour of the ratio for a specific length of the cavity as it is located at different $R$. According to the approximation estimator we considered, the longer $L$ cases are less accurate. However for all radii above $R=40$, the estimator \eqref{estimator1} remains less than $3\%$ above 1 even when the cavity proper length is $L=6$, which is the largest cavity length considered. 
 
We see from  Fig. \ref{ratiograph} that the larger the cavity, the greater the difference in excitation probability as the cavity is placed close to the horizon. As expected from the equivalence principle, very small cavities ($L=10^{-3}$) are virtually indistinguishable from the Rindlerian case --to see any distinction one would have to place the cavity much closer to the horizon than our present computational resolution admits. However departures from the Rindlerian case can be seen even for moderately small cavities ($L=0.3$), and these departures rapidly increase with cavity size provided one is within the vicinity of about 20 horizon radii.

A better comparison is given in Fig.\ref{ratiograph}b, in which we compute the ratio of the probabilities of  the two scenarios where the constant acceleration for the Rindlerian case is taken to be the  Schwarzschild acceleration in the middle of the cavity.  We see that the Schwarzschild case is consistently smaller than the Rindler case, with the discrepancy increasing with both increasing cavity size and closer proximity to the horizon.

\section{Conclusions}\label{concl}

We have introduced a cavity model in which we can find the transition probability of a Unruh-DeWitt detector in a curved spacetime without facing the difficulties of solving the wave equations, regularizations and defining the notion of vacuum state for different observers in a curved background. 
Our model works well where the interplay of the size of the cavity and strength of the curvature provides an environment in which a quasi-local approximation to the wave equation is valid.  

We studied the cavity setting in two scenarios: that of a freely falling detector in Schwarzschild and that of a uniformly accelerated detector in flat space-time crossing a stationary cavity. We found near-identical transition rates for both scenarios in the limit of small cavities, with increasing departures from this situation as the cavity size increases.

Consequently, in the limit of small cavity and large $R$ we are studying a particular case of the equivalence principle: the Schrwarzschild scenario is completely equivalent to a Rindler scenario   consisting of a cavity accelerating toward a stationary detector. We have shown that this scenario  coincides exactly with a setup where we have a uniformly accelerated detector crossing an inertial cavity.  Our work stands in contrast to that in which transition rates of detectors are computed for scalar fields in free space (for a review see \cite{Hu:2012jr}).

Comparing the Schwarzschild scenario to the Rindler scenario in which the uniform acceleration is taken to be the Schwarzschild acceleration (avarage of the field strength) in the middle of the cavity, we find that the latter case is consistently larger than the former, the discrepancy increasing with  both  closer proximity to the horizon and increasing cavity size. Our results show that the thermal radiation recorded by a detector in a Rindler space and the Hawking-like radiation that the detector observes in a Schwarzschild background approach the same quantity. 

For larger cavities, the quasi-local approximation breaks down for distances closer to the black hole where the curvature is large. To find the transition probability of the detector in this case, one must solve for the response function of the detector using the Wightman function in free space.

Note that the response of our detector is independent of the global vacuum choice  outside the cavity since, in our idealized setting,  the detector is `shieldedÕ by the cavity walls. For non-ideal cavities  the  outside field could leak inside the cavity, which would indeed have an effect on the response of the detector, particularly when the cavity is placed close to the event horizon. In those situations the choice of vacuum outside the cavity would become important. Considering these effects remains a subject of study for further research.

The methods we present in this paper should be applicable to a much broader class of situations.  Inclusion of mass is straightforward, and it would be interesting to study the effects of curvature relative to those previously obtained for uniform acceleration \cite{Dragan:2011zz}.   Of particular interest is to understand the effects of an ergoregion on transition probabilities.   Work on these areas is in progress.

%Two main difficulties: Complication in renormalization of the correlation function in curved spacetime. Sharp cut-off of the switching function at the singularity of the Wightman distribution. No general derivations in curved spacetimes, so far.

%\section{Outlook}\label{next}
%Same study for Kerr black hole. 
%Explore the circular trajectories in Schwarzschild and Kerr background on which detector is traveling. 
%Considering two entangled detectors and study the effect of curvature on their entanglement while one of them falling into the ergosphere of Kerr black hole.

\section{Acknowledgments}

The authors are thankful to B.L. Hu, A. Dragan, J. Louko, and T. Ralph for helpful and invaluable discussions. E.M-M. acknowledges the support of the Banting Postdoctoral Fellowship Programme. This work was supported in part by the Natural Sciences and Engineering Research Council of Canada.

\bibliography{cavity_refs}

\end{document}